\documentclass[pra,amsmath,amssymb,superscriptaddress,twocolumn,longbibliography,showpacs]{revtex4-1}
\usepackage{graphicx}
\usepackage{color}

\newcommand{\half}{\frac{1}{2}}

\begin{document}

\title{Counterfactual distribution of cat states}
\author{Akshata  Shenoy  H.}
\email{akshata@ece.iisc.ernet.in} 
\affiliation{Electrical  Communication Engg.  Dept.,  IISc, Bengaluru,
  India}   
\author{R. Srikanth}
\email{srik@poornaprajna.org}  \affiliation{Poornaprajna Institute
of Scientific Research,  Bengaluru,  India}

\begin{abstract}
In the counterfactual cryptography scheme  proposed by Noh (2009), the
sender Alice probabilistically transmits  classical information to the
receiver  Bob without  the physical  travel  of a  particle.  Here  we
generalize this idea to the distribution of quantum entanglement.  The
key insight is  to replace their classical input  choices with quantum
superpositions.  We further show that the scheme can be generalized to
counterfactually distribute multi-partite cat states.
\end{abstract}

\maketitle 

\section{Introduction}     
Counterfactual   quantum    communication   \cite{Joz98,MJ01}   allows
information to  be transmitted even  without the physical travel  of a
particle.   It is  based on  interaction-free measurements,  where one
tries to  detect a  quantum object  without directly  interrogating it
\cite{EV93}.  Ref.   \cite{KWH+95} proposed  a method to  increase the
efficiency  (of 50\%)  of the  counterfactual effect  in the  original
Elitzur-Vaidman  protocol  \cite{EV93}   towards  1  by  concatenating
unbalanced beam  splitters, each  of which effects  the transformation
$|0\rangle \rightarrow \cos\theta|0\rangle  + \sin\theta|1\rangle$ and
$|1\rangle  \rightarrow  -\sin\theta|0\rangle +  \cos\theta|1\rangle$.
This  constitutes  the quantum  Zeno  effect  (QZE), wherein  repeated
measurements or `interrogation' of an initial state freeze evolution.

Briefly, this works as follows: light in state $|0\rangle$ is incident
on  a  beam  splitter.   An  obstacle is  placed  in  the  output  arm
$|1\rangle$.   The  obstacle in  the  state  ``block'' eliminates  the
corresponding  amplitude in  the superposition,  whereas in  the state
``pass'', it does nothing.   Chaining the above interferometric action
leads to the evolution:
\begin{eqnarray}
|{\rm block}\rangle|0\rangle &\longrightarrow& \cos^L\theta
  |{\rm block}\rangle|0\rangle \nonumber \\
|{\rm pass}\rangle|0\rangle &\longrightarrow& 
  |{\rm pass}\rangle(\cos (L\theta)|0\rangle + \sin(L\theta)|1\rangle),
\label{eq:kwiat}
\end{eqnarray}
where  $L$   is  the  number   of  interferometric  cycles,   and  the
unnormalized state in  the first equation indicates  absorption at the
obstacle.   Setting  $\theta  = \frac{\pi}{2L}$,  let  $L  \rightarrow
\infty$.  The  result has full  counterfactual efficiency in  that the
obstacle set in block mode always  produces a distant detection in the
output $|0\rangle$ arm.

Counterfactuality  has   also  been   applied  to   both  cryptography
\cite{GS99} and  computation \cite{HRB+06,vai07}.  Since  the proposal
of  the  counterfactual  key   distribution  protocol  Noh-2009  (N09)
\cite{N09},  there has  been  an  upsurge of  interest  in this  area,
leading  to   contributions  devoted   to  improving   its  efficiency
\cite{SW10,ZLZ14},  to security  considerations under  various attacks
\cite{YLC+10,ZWT+12,ZWT12},   to    modified   communication   schemes
\cite{SLA+13,SSS1},    applications   \cite{ZGG+13,SSS14,Sal14}    and
experimental proposals  or realizations \cite{YLY+12,*BCD+12,*LJL+12}.

\section{Counterfactual quantum state transfer}

The  method  of Eq.   (\ref{eq:kwiat})  can  be  used for  direct  and
deterministic communication of classical bits  from Bob located at the
obstacle to Alice located at the interferometer output arms.  But as a
communication protocol, this method is not counterfactual for the case
``pass'' since  the photon encounters  no obstacle.  To make  both bit
values counterfactual, \cite{SLA+13} proposes the chained quantum Zeno
effect (CQZE) set-up, which nests the above $L$-chained interferometer
within a chain  of $M$ ``outer'' unbalanced beam splitters,  in such a
way  that  at each  inner  cycle,  the  outcome arm  corresponding  to
``pass''  would  lead  to  detection  at  a  detector  $\Delta$.   One
sequentially   evolves   the  state   through   each   of  the   outer
interferometric  cycles,   which  gives   a  recursion   relation  for
amplitudes  for each  cycle, analogous  to Eq.   (\ref{eq:cgate}).  By
suitable choice of $L$ and $M$, one can obtain direct communication of
bits from Alice  to Bob. This is argued to  be fully counterfactual in
the sense that Bob's both ``pass'' and ``block'' choices correspond to
blocking    actions.     (Regarding     this    interpretation,    see
Refs. \cite{vai14, *SLA+14, *Gis13, *Vaid14}.)

Now, the method of Eq. (\ref{eq:kwiat})  can be readily adapted to one
for direct  communication of  \textit{qubits} (rather than  bits) from
Bob to Alice by letting the  obstacle to be in the superposition state
$\alpha|{\rm     pass}\rangle      +     \beta|{\rm     block}\rangle$
\cite{S14a,GCC+14}.   An initial  state  $(\alpha|{\rm pass}\rangle  +
\beta|{\rm block}\rangle)  \otimes |0\rangle$ of the  obstacle and the
particle   evolves   to   $(\alpha|{\rm  pass}\rangle   +   \beta|{\rm
  block}\rangle) \otimes  (\cos\theta|0\rangle + \sin\theta|1\rangle)$
after the particle passes through  the first beam splitter.  Following
the particle's interaction with the obstacle placed in arm $|1\rangle$
after      the     beam      splitter,      the     state      becomes
$\sqrt{1-\beta^2\sin^2(\theta)}        [\alpha|{\rm       pass}\rangle
  (\cos\theta|0\rangle    +    \sin\theta|1\rangle)    +    \beta|{\rm
    block}\rangle |0\rangle]$.   Here the global pre-factor,  which is
the square root of the particle's survival probability, comes from the
fact that  $\beta^2\sin^2(\theta)$ is the probability  that the photon
is absorbed at the obstacle.  Just after the second beam splitter, but
before     encountering     the     obstacle,     the     state     is
$\sqrt{1-\beta^2\sin^2(\theta)}        [\alpha|{\rm       pass}\rangle
  (\cos(2\theta)|0\rangle   +  \sin(2\theta)|1\rangle)   +  \beta|{\rm
    block}\rangle         (\cos\theta|0\rangle+\sin\theta|1\rangle)]$.
Proceeding  thus, just  after  the application  of  $L$ beam  splitter
passages, the state transforms as:
\begin{eqnarray}
&&(\alpha|{\rm  pass}\rangle +  \beta|{\rm
  block}\rangle)|0\rangle 
\rightarrow \left[1-\beta^2\sin^2(\theta)\right]^{(L-1)/2} \times \nonumber\\
&&\hspace{0.5cm} 
  \left[\alpha|{\rm pass}\rangle(\cos (L\theta)|0\rangle + \sin(L\theta)|1\rangle) \right. \nonumber \\
 &&\hspace{0.5cm} + \left.
\beta |{\rm block}\rangle(\cos\theta|0\rangle + \sin\theta|1\rangle)\right]
\nonumber \\
&&\hspace{0.5cm}
\rightarrow \alpha|{\rm pass}\rangle|1\rangle
+ \beta|{\rm block}\rangle|0\rangle
\label{eq:cgate}
\end{eqnarray}
where the right-arrow in the  last equation indicates taking the limit
$L \rightarrow \infty$ with  $\theta=\frac{\pi}{2L}$.  (Note that this
limit restores the normalization.)

If the  $L$-cycle is nested in  an $M$-cycle as described  above, then
analogous to the superposition (\ref{eq:cgate}), the final state after
passing through the $M$ outer cycles, by adjusting $L$ and $M$, can be
brought to the form:
\begin{equation}
|\Psi_{\rm fin}\rangle \approx \alpha|{\rm pass}\rangle|x\rangle
-\beta|{\rm block}\rangle|y\rangle,
\label{eq:fin-}
\end{equation}
where $x, y$ label the output  arms of the $M$th outer interferometer.
Note  that the  entangled  state $|\Psi_{\rm  fin}\rangle$ depends  on
Bob's   initial   state   $\alpha|{\rm  block}\rangle   +   \beta|{\rm
  pass}\rangle$. With  a Hadamard  and 1-bit  classical communication,
Alice can recover Bob's state  in the $\{|x\rangle, |y\rangle\}$ basis
(see later below).   This forms the essence of the  protocol of Guo et
al.  \cite{GCC+14} for quantum state transfer.  It requires only 1 bit
for  deterministic quantum  state  transfer,  unlike standard  quantum
teleportation,  which requires  2 bits,  because teleportation  uses a
standard Bell  state as the  entanglement resource, whereas  the above
protocol  uses the  state-dependent entanglement  (\ref{eq:fin-}).  We
note  that  even without  the  classical  communication, the  protocol
succeeds  half the  time, such  that the  fidelity of  the transmitted
mixed   state  is   1   for  polar   qubits,   falling  gradually   as
$|\cos(\theta)|$  towards  0 for  equatorial  $(\theta=\frac{\pi}{2})$
qubits  of  the Bloch  sphere  in  the $\{|{\rm  block}\rangle,  |{\rm
  pass}\rangle\}$ basis.
  
That the counterfactual  quantum protocol to transmit  a classical bit
\cite{SLA+13} is direct and deterministic suggests that a protocol for
quantum state transfer built on top of  it can also be that way.  Ref.
\cite{S14a} proposes one way to do this: a polarization-based ``dual''
CQZE scheme, which essentially first entangles Alice's and Bob's input
during a ``counterfactual CNOT operation'', and then disentangles them
using a  second counterfactual CNOT  in such  a way that  the combined
action swaps  Alice's state $|0\rangle$ with  Bob's state $\alpha|{\rm
  pass}\rangle  + \beta|{\rm  block}\rangle$ deterministically.   Note
that the  swap of two  generic qubit  states requires three  CNOTs, in
constrast  to the  two CNOTs  that suffice  for the  states considered
here.

In the present work, we propose to use N09 instead of the CQZE system,
as the  basis for  counterfactual distribution of  entanglement.  Like
the  state  (\ref{eq:fin-}),  this  entanglement will  depend  on  the
initial  superposition states  of  Bob's obstacle  and Alice's  input.
Unlike them, however, in our  case the state-dependent entanglement is
generated    \textit{probabilistically}.     Once   generated,    this
entanglement can be used for deterministic quantum state transfer with
a 1-bit  communication, as  in the  Guo et  al. protocol.   Though our
method has a lower yield than the methods of Refs. \cite{GCC+14,S14a},
still our experimental  set-up is simpler in that we  do not require a
large    chaining   of    beam-splitters.    Further,    the   schemes
\cite{S14a,GCC+14} pertain  to the bipartite case,  whereas our method
is   generalized  to   one   that   can  counterfactually   distribute
multipartite quantum states, in particular $(N+1)$-partite cat states,
where $N$,  an integer greater  than 1, is  the number of  players who
initiate the protocol.

The  remaining  article  is  structured  as  follows.   After  briefly
describing  the N09  protocol  in Section  \ref{sec:N09},  we show  in
Section \ref{sec:new}  how to modify  it for the purpose  of bipartite
entanglement distribution,  essentially by replacing bits  $a$ and $b$
by  qubits,  and  correspondingly   replacing  $R_a$  and  $R_b$,  the
classical  random number  generators, by  their quantum  counterparts,
which  can  generate   superposition  states.   Section  \ref{sec:qst}
discusses using the counterfactually  distributed entanglement for the
purpose of quantum state transfer,  comparing our method with those of
counterfactual  schemes  of  \cite{GCC+14,S14a} and  standard  quantum
teleportation.   In Section  \ref{sec:xcat}  we  extend the  bipartite
scheme to that of generating multipartite cat states, by introducing a
quantum network  with a star  topology.  In Section  \ref{sec:exp}, we
discuss  candidate  physical  systems  for  practical  implementation.
Finally  we  discuss some  implications  of  our scheme  for  physical
interpretation, the extension of the scheme of Ref.  \cite{GCC+14} for
the  counterfactual  distribution  of   cat  states,  and  offer  some
conclusions in Section \ref{sec:conclu}.

\section{Counterfactual        communication       of        classical
  bits: the N09 protocol \label{sec:N09}} 

To set the  background, we briefly describe N09 as  follows. The basic
set-up of the N09 is similar to the one shown in Fig.  \ref{fig:cat1}.
The sender (Alice) and the receiver  (Bob) are connected to each other
through  arm  $B$ of  a  Michelson  interferometer.   The arm  $A$  is
internal to Alice's station and a photon travelling along this path is
reflected using a Faraday mirror  (FM).  Alice's station also consists
of a single photon source  ($S$) which prepares polarization states in
the vertical ($V$) or horizontal  ($H$) direction, based on the output
$a$ of a random number generator  $R_a$.  Bob's module $Q$ consists of
detector  $D_B$ and  an  FM which  absorbs or  reflects  photons in  a
polarization-dependent way based on the state of a switch.  The switch
state is $P$ (``pass $V$ and block $H$'') or $B$ (``block $V$ and pass
$H$''), as determined  by the output $b$ of a  random number generator
$R_b$.  The protocol runs as follows:
\begin{enumerate}
\item Depending  on the random  bit $a$, Alice prepares  single photon
  states randomly  in $V$ or  $H$ polarization, and transmits  them to
  Bob, who applies $P$ or $B$ actions randomly based on the random bit
  $b$.
\item The possibilities of Alice's and Bob's random switching actions,
  and the corresponding conditional probabilities, are:
\begin{equation}
\begin{array}{c||c|c|c}
\hline
\textrm{(Alice,~Bob)} & D_1 & D_2 & D_B \\
\hline 
(V,P)~\textrm{or}~(H,B) & 0 & 1 & 0 \\
\hline
(V,B)~\textrm{or}~(H,P) & RT & R^2 & T\\
\hline
\end{array}
\label{eq:Nohprop}
\end{equation} 
In  the  first  case,  deterministic detection  of  $D_2$  because  of
constructive interference between the two reflected photons.  Here $R$
and  $T$  are  the   coefficients  of  reflectance  and  transmittance
respectively of the beamsplitter (BS), and satisfy $R+T=1$.
\item After the protocol run, Alice announces instances where she made
  $D_1$ detections, and use the polarization  ($V$ or $H$) as a secret
  bit.  The $D_2$ detections are used for security check.
\end{enumerate}
The case of $D_1$ is counterfactual in the sense that the particle did
not physically travel to Bob's  station, and his blocking action leads
to a  remote detection by  Alice.  The counterfactual yield  is $RT/2$
and, in the noiseless case, would represent the positive key rate.

\section{Counterfactual bipartite entanglement generation:
A new scheme \label{sec:new}} 

\begin{figure}
\begin{center}
\includegraphics[width=0.4\textwidth]{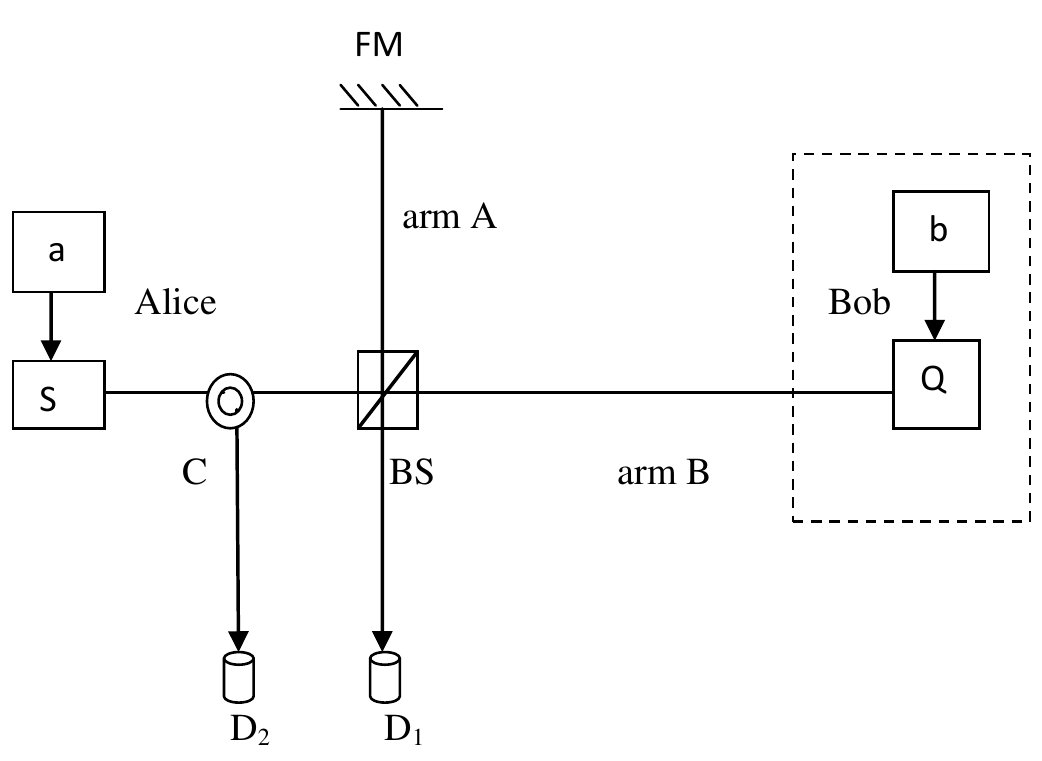}
\caption{Basic  schematic  for N09,  where  $a$  and $b$  take  random
  classical  bit   values.   For  its  adaptation   to  counterfactual
  generation of  bipartite entanglement, $a$  and $b$ are  qubits that
  can  exist  in  arbitrary   superposition  states.   $C$  denotes  a
  circulator.   The   switch  $Q$  implements   pass-$V$-block-$H$  or
  pass-$H$-block-$V$  on the  incoming photon,  depending on  $b$.  To
  distribute  entanglement  counterfactually,  $Q$ is  entangled  with
  Bob's input $b$.}
\label{fig:cat1}
\end{center}
\end{figure}

We shall now  adapt N09 to distribute quantum  entanglement.  We begin
with  the  demonstration  of  our idea  to  counterfactually  generate
bi-partite entanglement.  The basic idea  can be indicated again using
Figure \ref{fig:cat1}.   The key new  element is that $R_a$  and $R_b$
are now \textit{quantum} random  number generators that produce qubits
$a$ and $b$.
Let qubit $b$ be given by $|\psi\rangle_b = \alpha |P\rangle_b + \beta
|B\rangle_b$  and  $a$ by  $|\phi\rangle_a  =  \mu |V\rangle_a  +  \nu
|H\rangle_a$.  Then,  the initial  joint states  of the  $R_a$, $R_b$,
Bob's detector and the optical field can be written as,
\begin{eqnarray}
|\Psi_0\rangle &=& |\phi\rangle_a|0\rangle_{D_B}|\psi\rangle_b|\Psi\rangle_{AB}
\\ 
&=& 
(\mu|V\rangle_a + \nu|H\rangle_a)
|0\rangle_{D_B}
(\alpha|P\rangle_b + \beta|B\rangle_b)
|0\rangle_{A}|0\rangle_B,\nonumber
\label{eq:Psi0}
\end{eqnarray}
where  $|\Psi_0\rangle_{AB}  \equiv |0\rangle_A|0\rangle_{B}$  is  the
initial vacuum state of the channel, represented by the interferometer
arms.

Alice's random number generator  $R_a$ now determines the polarization
state of the  photon.  Under interaction with $R_a$ and  BS, the state
$|\Psi_0\rangle$ evolves to:
\begin{eqnarray}
|\Psi_1\rangle &=& (\alpha|P\rangle_b + \beta|B\rangle_b)|0\rangle_{D_B} \otimes \nonumber \\
  && \left(\mu|V\rangle_a\left[i\sqrt{R}|V\rangle_A|0\rangle_B + \sqrt{T}|0\rangle_A|V\rangle_B\right]
 \nonumber\right. \\
  &+&  \left.    \nu|H\rangle_a\left[i\sqrt{R}|H\rangle_A|0\rangle_B + \sqrt{T}|0\rangle_A|H\rangle_B\right]
  \right)
\label{eq:Psi1}
\end{eqnarray}
Let  $|0\rangle_{D_B}$  and  $|Y\rangle_{D_B}$ represent  the  initial
state of Bob's detector, and the state of his detector after detecting
a photon.  We represent the switch action on Bob's side thus:
\begin{eqnarray}
|P\rangle_b|V\rangle_B|0\rangle_{D_B} &\rightarrow&
|P\rangle_b|V\rangle_B|0\rangle_{D_B} \nonumber \\
|P\rangle_b|H\rangle_B|0\rangle_{D_B} &\rightarrow&
|P\rangle_b|0\rangle_B|Y\rangle_{D_B} \nonumber \\
|B\rangle_b|V\rangle_B|0\rangle_{D_B} &\rightarrow&
|B\rangle_b|0\rangle_B|Y\rangle_{D_B} \nonumber \\
|B\rangle_b|H\rangle_B|0\rangle_{D_B} &\rightarrow&
|B\rangle_b|H\rangle_B|0\rangle_{D_B}
\label{eq:Bobdet}
\end{eqnarray}
Post-selected  on  Bob's  detector being  found  in  the  state
  $|0\rangle_{D_B}$,  the  optical  field  in  state  $|\Psi_1\rangle$
  evolves under (\ref{eq:Bobdet}) to the un-normalized state
\begin{eqnarray}
|\Psi_2\rangle &=& 
\alpha|P\rangle_b \left(\mu|V\rangle_a\left[i\sqrt{R}|V\rangle_A|0\rangle_B + \sqrt{T}|0\rangle_A|V\rangle_B\right] 
\right.
\nonumber\\
 &&~+  \left. \nu|H\rangle_a\left[i\sqrt{R}|H\rangle_A|0\rangle_B\right]\right) 
   \nonumber \\
&+&
\beta|B\rangle_b\left(\mu|V\rangle_a\left[i\sqrt{R}|V\rangle_A|0\rangle_B 
\right]\right.\nonumber\\ 
&&~+ \left. \nu|H\rangle_a\left[i\sqrt{R}|H\rangle_A|0\rangle_B 
   + \sqrt{T}|0\rangle_A|H\rangle_B\right]\right),
\label{eq:Psi2}
\end{eqnarray}
with  probability $P(D_B)$  for Bob's  detector being  found in  state
$|Y\rangle_{D_B}$  being  
\begin{equation}
P(D_B)=1-\left|\langle\Psi_2|\Psi_2\rangle\right|^2=(|\alpha\nu|^2+|\beta\mu|^2)T.
\label{eq:PDB}
\end{equation}
In  the  last  phase,  the  light re-enters  Alice's  system.   It  is
convenient  to represent  the BS  transformation in  the return  path,
together with the reflection on Bob's side, by:
\begin{eqnarray}
|0\rangle_A|x\rangle_B &\longrightarrow&
\sqrt{R}|D_1^x\rangle + i\sqrt{T}|D_2^x\rangle \nonumber \\
|x\rangle_A|0\rangle_B &\longrightarrow&
i\sqrt{T}|D_1^x\rangle + \sqrt{R}|D_2^x\rangle,
\label{eq:beamsplitter}
\end{eqnarray}
where $x  = V,H$,  and $|D_j^x\rangle$ denotes  the state  of detector
$D_j$ when it detects a photon of polarization $x$.

Thereby the post-selected state $|\Psi_2\rangle$ evolves to
$|\Psi_3\rangle \equiv |\Psi_3^\prime\rangle|0\rangle_{D_B}$, where
\begin{widetext}
\begin{eqnarray}
|\Psi_3^\prime\rangle &=&
\alpha|P\rangle_b \left(\mu|V\rangle_a
|D_2^V\rangle
+\nu|H\rangle_a
\left[i\sqrt{R}(i\sqrt{T}|D_1^H\rangle + 
\sqrt{R}|D_2^H\rangle) 
\right]\right) \nonumber \\
&+& 
\beta|B\rangle_b\left(\mu|V\rangle_a\left[i\sqrt{R}
(i\sqrt{T}|D_1^V\rangle + \sqrt{R}|D_2^V\rangle) 
\right]
+\nu|H\rangle_a|D_2^H\rangle\right). \label{cat3}
\end{eqnarray}
\end{widetext}
If the superpositions in $a$ and  $b$ are replaced by their projective
measurements prior to the photon entering the interferometer, then the
present  protocol  reduces  to  N09,  in  particular  reproducing  the
conditional  probabilities (\ref{eq:Nohprop}).  For example,  from Eq.
(\ref{eq:PDB}), we  have $P(D_B)=(|\alpha\nu|^2+|\beta\mu|^2)T$.  From
Bayesian        analysis        we        have        $P(D_B)        =
P(D_B|H_aP_b)P(H_aP_b)+P(D_B|V_aB_b)  P(V_aB_b)$.   Now  $P(H_aP_b)  =
P(H_a)P(P_b)=|\alpha\nu|^2$ and $P(V_aB_b)=P(V_a)P(B_b)=|\beta\mu|^2$,
where  $P(H_a)$  etc.,  are  probabilities to  obtain  the  designated
outcome  under  projective  measurement of  the  corresponding  device
($R_a$ or  $R_b$).  Equating  these two  expressions for  $P(D_B)$ and
noting that this holds true for arbitrary protocol parameters, we find
that    $P(D_B|H_aP_B)    =     P(D_B|V_aB_b)=T$,    as    given    in
Eq. (\ref{eq:Nohprop}).

Now, conditioned  on $D_1$  being detected  (which corresponds  to the
projector  $\sum_{H,V}  |D_1^x\rangle\langle  D_1^x|$),  we  have  the
counterfactual situation with the resulting (unnormalized) state being
\begin{equation}
|\xi\rangle   =    \sqrt{RT}\left(\alpha\nu|H\rangle_a|P\rangle_b   +
\beta\mu|V\rangle_a|B\rangle_b\right),
\label{eq:fin}
\end{equation} 
where  the counterfactual  yield  is seen  to  be 
\begin{equation}
P(D_1) = RT(|\alpha\nu|^2  + |\beta\mu|^2).
\label{eq:PD1}
\end{equation}  
We note  that the beam  splitter function $\sqrt{RT}  = \sqrt{R(1-R)}$
does   not   determine   the    degree   of   entanglement   generated
counterfactually,  which depends  only  on  the initial  superposition
states  held by  Alice  and Bob.   A $D_B$  detection  results in  the
entangled            (unnormalized)           state            $T\left
(\alpha\nu|H\rangle_a|P\rangle_b         +         \beta\mu|V\rangle_a
|B\rangle_b\right)$,  which differs  from  state  (\ref{eq:fin}) by  a
constant  factor.  However  it  is not  counterfactual.  Similarly,  a
$D_2$ detection, which happens with probability
\begin{equation}
P(D_2) = |\alpha|^2(|\mu|^2 + |\nu|^2R^2) + 
|\beta|^2(|\nu|^2 + |\mu|^2R^2),
\label{eq:PD2}
\end{equation}
non-counterfactually produces  the state $\alpha(\mu|V\rangle_a  + \nu
R|H\rangle_a)|P\rangle_b      +     \beta(i\mu      R|V\rangle_a     +
\nu|H\rangle_a)|B\rangle_b$, which is in general entangled.

Intuitively,  the  above  scheme  can  be  thought  of  as  tripartite
entanglement  being established  between  the  photon, Alice's  device
$R_a$ and Bob's device $R_b$, after  which the photon is projected out
in a suitable basis, leaving $R_a$ and $R_b$ entangled.

\section{Counterfactual quantum information transfer
in the  new scheme  \label{sec:qst}} 

By means  of the state-dependent  entanglement so obtained,  a quantum
state transfer can be implemented either from Alice to Bob or from Bob
to Alice,  with a 1-bit communication.   The sender puts the  input in
the required unknown state, while  the receiver prepares her/his input
in an equal-weighted superposition in the default basis.  For example,
suppose that it is desired to transfer Alice's state $|\phi\rangle_a =
\mu|V\rangle_a + \nu|H\rangle_a$ to Bob.  Bob prepares his input state
with $\alpha=\beta=\frac{1}{\sqrt{2}}$.  From  Eq.  (\ref{eq:fin}), we
see  that  Alice and  Bob  (probabilistically)  will share  the  state
$\nu|H\rangle_a|P\rangle_b  + \mu|V\rangle_a|B\rangle_b$.   Applying a
Hadamard on Alice's qubit in the $\{V,H\}$ basis results in the state
$$
\frac{|V\rangle_a}{\sqrt{2}}(\nu|P\rangle_b - \mu|B\rangle_b) +
\frac{|H\rangle_a}{\sqrt{2}}(\nu|P\rangle_b + \mu|B\rangle_b).
$$ Measuring  in the $\{V,H\}$  basis, and depending on  whether Alice
finds outcome  $V$ or $H$,  she communicates  a 1-bit message  to Bob,
which instructs  him to apply  the Pauli operation  $Z$ or $I$  in the
$\{P,  B\}$ basis.  Bob is  then  left with  the reconstructed  state:
$|\phi\rangle_b \equiv \mu|B\rangle_b + \nu|P\rangle_b$.

This   situation   is   analogous   to  that   encountered   in   Ref.
\cite{GCC+14}, except  that there the state-dependent  entanglement is
deterministically generated (in the asymptotic  limit of the number of
beam   splitters),   whereas   in    our   case,   it   is   generated
probabilistically,   with   probability  $P(D_1)\equiv\frac{RT}{2}   =
\frac{R(1-R)}{2}$, as  follows from  Eq. (\ref{eq:PD1}).   Denoting by
the term  $C_q$ the number of  photons consumed (or runs  required) on
average per  counterfactual event, we  have $C_q =  \frac{1}{P(D_1)} =
\frac{2}{R(1-R)}$. The minimum value  of $C_q$ is $C_q^{\rm min}\equiv
8$ photons (or runs), which is attained at $R=\half$.  In other words,
a balanced  beam splitter  minimizes the  quantum resource  used.  The
situation  is slightly  different for  the average  \textit{classical}
communication required, as shown below.

Assuming the communication  channel to be noiseless, and  the light to
consist  of   single-photons,  the  average  communication   cost  for
generating an entangled pair is:
\begin{equation}
C = h\left(\frac{R(1-R)}{1+R}\right)\frac{1+R}{R(1-R)},
\label{eq:CC}
\end{equation}
where $h(x)\equiv-x\log_2(x)-(1-x)\log_2(1-x)$  is the  Shannon binary
entropy.   From Eq.   (\ref{eq:CC}), we  see that  cost $C$  increases
without bound  as $R  \rightarrow0$ or  $R\rightarrow1$.  This  can be
anticipated, referring to Eq.   (\ref{eq:Nohprop}), where we find that
the number of  counterfactual ($D_1$) events tend to  vanish in either
limit, thereby  necessitating a  large communication cost.   Note that
since each run requires at most  one bit communication, therefore as a
thumb rule, cost $C$ in (\ref{eq:CC})  cannot be larger than $C_q$, at
a given $R$.

To derive  (\ref{eq:CC}), suppose that  $n$ runs have  been conducted,
and Bob's input state is chosen with $\alpha=\beta=\frac{1}{\sqrt{2}}$
in each run.  Then the  probabilities for the different outcomes $D_1,
D_2$  and  $D_B$  are  just   the  conditional  probabilities  in  Eq.
(\ref{eq:Nohprop}) with  a factor $\frac{1}{2}$, independent  of $\mu$
and $\nu$,  as follows  from Eqs.  (\ref{eq:PDB}),  (\ref{eq:PD1}) and
(\ref{eq:PD2}).   For  sufficiently  large $n$,  about  $\frac{nT}{2}$
detection events $D_B$ occur, during which Alice has no detection, but
obviously does not need to  communicate this information to Bob.  Thus
on average,  only on the  remaining $  n_B \equiv nP(\neg  D_B) \equiv
n\left(1-\frac{T}{2}\right) =  \frac{n(1+R)}{2}$ of events  does Alice
need  to communicate  to  Bob about  whether a  $D_1$  or $D_2$  event
happened.

Detections $D_1$ and $D_2$ occur  among these events, with probability
$P(D_1)\equiv    RT/2    =   \frac{R(1-R)}{2}$    and    $P(D_2)\equiv
\frac{1+R^2}{2}$, respectively, as seen  from Eqs.  (\ref{eq:PD1}) and
(\ref{eq:PD2}).  Therefore the  probability of a $D_1$  or $D_2$ event
in  the  outcome space  of  non-$D_B$  events is  $p^\prime(D_1)\equiv
P(D_1)/P(\neg   D_B)   =   \frac{R(1-R)}{1+R}$   and   $p^\prime(D_2)=
1-p^\prime(D_1)  = \frac{1+R^2}{1+R}$,  respectively.  Therefore,  the
bit string about $n_B$ long,  corresponding to events when Alice needs
to  communicate, can  be  compressed (by  Shannon's noiseless  channel
coding theorem \cite{NC00}) to  $n_B^\ast \equiv h[p^\prime(D_1)] n_B$
bits.   On  average,  these  events  resulted  in  the  counterfactual
distribution    of    $n_1    \equiv    n    P(D_1)=\frac{n}{2}R(1-R)$
state-dependent entangled states.  The  ratio $n_B^\ast/n_1$ gives the
required  expression  for  the average  classical  communication  cost
(\ref{eq:CC}).

To minimize cost  $C$, note that the rhs in  Eq. (\ref{eq:CC}) has the
form $h(\xi)/\xi$, which  falls monotonically as $\xi$ goes  from 0 to
1.   Now the  argument  of $h(\cdot)$  in  Eq.  (\ref{eq:CC}),  namely
$\xi\equiv  R(1-R)/(1+R)$, takes  values  only in  the range  $[0,1]$.
Thus the required minimum for $C$ is obtained by minimizing $\xi$ as a
function  of  $R$, which  is  seen  to occur  at  $R=\sqrt{2}-1\approx
0.414$,  with  the  corresponding  average  communication  cost  being
$C^{\rm  min} \equiv  C|_{R=\sqrt{2}-1} \approx  3.85$ bits.   We note
that  $C^{\rm min}  \le C^{\rm  min}_q$, and  that the  latter minimum
occurs at  $R=\half$.  For the  case of  a balanced beam  splitter, we
find $C|_{R=0.5}\approx  3.9 >  C^{\rm min}$.   The minimum  number of
bits  of  classical  communication   required  on  average  using  our
counterfactual  teleportation  is  $C^{\rm min}+1\approx  4.85$  bits.
This is larger than the 2 bits sufficient (and necessary) for standard
quantum teleportation, 1 bit for  the \cite{GCC+14} protocol and 0 bit
for the protocol of \cite{S14a}.  Of course, these three schemes being
deterministic, $C_q=1$ for each of them.

\section{Extension to cat states\label{sec:xcat}}  

We  now  show  how  to  extend  the  above  method  of  counterfactual
generation of bipartite  entanglement to $(N+1)$-partite entanglement.
It  will be  convenient  here  to regard  the  `quantized' version  of
(\ref{eq:Nohprop}) as the  action of a propagator.   Then the `partial
propagator' relevant here has the action:
\begin{eqnarray}
|P\rangle_b |H\rangle_a  |\Psi\rangle_{AB}&\rightarrow& 
\sqrt{RT}|P\rangle_b |H\rangle_a|D_1^H\rangle \nonumber \\ 
|B\rangle_b |V\rangle_a  |\Psi\rangle_{AB}&\rightarrow& 
\sqrt{RT}|B\rangle_b |V\rangle_a |D_1^V\rangle,
\label{eq:recipe}
\end{eqnarray}
where by `partial propagator' we mean  that the evolved final state is
post-selected only on $D_1$ detections.

\begin{figure}
\begin{center}
\includegraphics[width=0.3\textwidth]{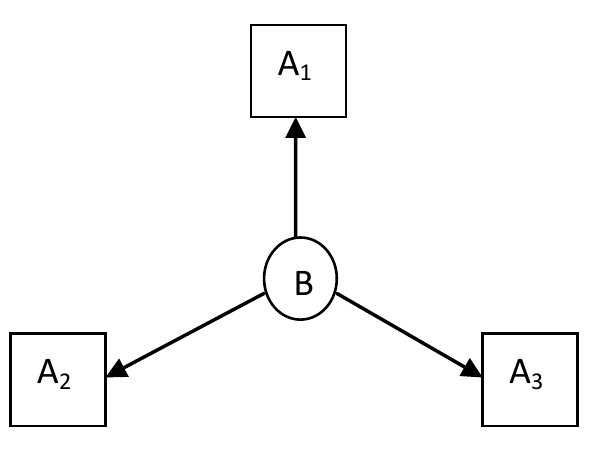}
\caption{Multipartite  counterfactual  entanglement  distribution:  We
  consider a star topology, with a  single Bob at the network hub, who
  shares the set-up  of Figure \ref{fig:cat1} with each  of $N$ (here:
  $N=3$) Alices. Bob's  quantum choice must be jointly  applied to all
  $N$ channels on his side.}
\label{fig:cat}
\end{center}
\end{figure}

Now  we  can  generalize  the   set-up  of  Figure  \ref{fig:cat1}  to
counterfactually distribute  an $(N+1)$-cat state, as  shown in Figure
\ref{fig:cat},  which illustrates  the  network  for the  distribution
among four  parties. Typically,  it is described  by a  star topology,
with Bob  at the network  hub and the  remaining players (who  we call
Alice$_1$, Alice$_2$, et  al., for convenience) situated  at the spoke
nodes.   Bob  is   connected  according  to  the   circuit  of  Figure
\ref{fig:cat1}  with  each  Alice$_j$ \textit{separately}.   The  only
requirement is that  Bob's quantum device $R_b$ makes  a common choice
for all Alices' particles.

For instance, considering the case of  $N=2$, the initial state of the
two Alices  and Bob,  given by  $ |\Psi_4\rangle  = \bigotimes_{j=1}^2
(\mu_j|V\rangle +  \nu_j|H\rangle)_{a_j} \otimes  (\alpha|P\rangle_b +
\beta|B\rangle_b) |\Psi\rangle_{A_1B_1} |\Psi\rangle_{A_2B_2}, $ where
$|\psi\rangle_{A_1B_1}  (|\psi\rangle_{A_2B_2})$ indicate  the optical
channel  between  Alice$_1$ (Alice$_2$)  and  Bob,  evolves under  the
partial propagator (\ref{eq:recipe}) to
\begin{equation}
|\Psi_{\rm cat}\rangle_{a_1a_2b} = RT\left(\mu_1\mu_2\beta|VVB\rangle +
\nu_1\nu_2\alpha|HHP\rangle\right)_{a_1a_2b},
\end{equation}
with  the  counterfactual  yield being  $(RT)^2(|\mu_1\mu_2\beta|^2  +
|\nu_1\nu_2\alpha|^2)$.

It is  straightforward to  extend the above  exercise to  $N$ Alice's,
thereby producing a $(N+1)$-cat state between these Alices and Bob.
The initial state is taken to be
\begin{equation}
\bigotimes_{j=1}^N (\mu_j|V\rangle_j + \nu_j|H\rangle_j)_{a_j}
\otimes
(\alpha|P\rangle +
\beta|B\rangle)_b,
\end{equation}
which evolves under the  evolution (\ref{eq:recipe}), corresponding to
post-selection on all Alices detecting $D_1$, to the state:
\begin{equation}
|\Psi_7\rangle =
(RT)^{N/2}\alpha\Pi_\mu|V\rangle_a^{\otimes     N}|B\rangle_b     +
\beta\Pi_\nu|H\rangle_a^{\otimes N} |P\rangle_b,
\label{eq:catN}
\end{equation}
where  $\Pi_\mu  \equiv  \mu_1\mu_2\cdots\mu_N$  and  $\Pi_\nu  \equiv
\nu_1\nu_2\cdots\nu_N$.

By setting $\alpha = \beta =  \frac{1}{\sqrt{2}}$ and $\mu_j = \nu_j =
\frac{1}{\sqrt{2}}$, we  obtain a maximally entangled  cat state, with
yield   $\left(\frac{RT}{2}\right)^N$,   indicating   an   exponential
fall-off.

\section{Experimental implementation\label{sec:exp}}

Recent experimental advances in  quantum information processing and in
the demonstration of quantum coherence  in mesoscopic systems has been
impressive, examples  being trapped  ion systems \cite{W13}  and large
bio-molecules \cite{G+11}. The key experimental challenge to implement
our proposal is putting Bob's  obstacle in the superposition of ``pass
H, block  V'' and ``pass V,  block H'', which is  a highly non-trivial
requirement.  

Considerable experimental simplification happens  in our scheme, while
at  the   same  time   preserving  essentially   the  same   idea  for
counterfactual  distribution   of  entanglement,  if  we   employ  the
semi-counterfactual quantum  key distribution  (\mbox{ScQKD}) protocol
\cite{SSS1},  rather  than   N09,  as  the  point   of  departure  for
counterfactual  communication from  classical to  quantum information.
Moreover,  a Mach-Zehnder  interferometer  version,  rather than  that
using  the  Michelson  interferometer,  may  be  advantageous  for  an
experimental demonstration, as is done in the experimental realization
of N09 \cite{BCD+12}.

For the  present purpose, the  ScQKD scheme  is similar to  N09 except
that secret bits are encoded in  terms of plain ``pass'' and ``block''
actions, rather than the  polarization-specific blockade actions.  The
Michelson interferometer  in Figure  \ref{fig:cat1} remains  the same,
except  that   the  fixed  mirror  at   the  top  is  replaced   by  a
polarization-independent  pass/block set-up  similar  to Bob's.   This
set-up now forms Alice's module.  Counterfactual bits are generated in
$D_1$ detections and  happen only when precisely one of  Alice and Bob
applies ``pass'' and the other ``block''.  To be precise, the bits are
counterfactual only with respect to  the person applying the blockade.
Thus it is counterfactual only on part of the runs with respect to the
distant agent,  Bob, hence  the name  \textit{semi-}counterfactual. By
replacing Alice's and Bob's classical choices by superpositions of the
type $\alpha_j|{\rm pass}\rangle  + \beta_j|{\rm block}\rangle$ ($j=A,
B$), one counterfactually and probabilistically generates entanglement
of the (unnormalized) form
\begin{equation}
\alpha_A\beta_B|{\rm pass}\rangle_A|{\rm block}\rangle_B
+ 
\beta_A\alpha_B|{\rm block}\rangle_A|{\rm pass}\rangle_B,
\label{eq:semic}
\end{equation}
where the first and second registers refer to Alice's and Bob's
systems.

We note  that in ScQKD, both  bits are counterfactual with  respect to
one  of  Alice  and  Bob.   Moreover,  Alice  and  Bob  are  spatially
separated. Both these features are relaxed in the Kwiat et al.  scheme
\cite{KWM+99} (where only the  ``block'' option is counterfactual with
respect to Alice or Bob), though at  the price of requiring a chain of
beam splitter actions.   Thus, an alternative, and in  some ways, even
simpler   first   step    towards   experimental   implementation   of
counterfactual   distribution    of   entanglement   would    be   the
semi-counterfactual   generation   of   entanglement   of   the   type
(\ref{eq:cgate}).

We suggest two broad ways to experimentally achieve such superposition
states (\ref{eq:semic})  of the  obstacle: one  using \textit{spatial}
Schr\"odinger  cat  states and  the  other  using  a special  kind  of
electromagnetically induced transparency (EIT) systems.

In the  first way, Bob's system  is put in the  superposition of being
present  \textit{and}  absent  in   the  communication  channel.   The
presence of the obstacle would  correspond to the ``block'' action and
the  absence to  the  ``pass''  action on  the  photon.  Trapped  ions
\cite{W13} are possible candidate systems here.

The second  way to  realize the  quantum obstacle  is as  a mesoscopic
system  placed in  a quantum  superposition of  being transparent  and
opaque to photons  passing through the channel.   A possible candidate
here is  of a cloud of  Rydberg atom confined in  a trapping potential
localized within  the blockade  radius \cite{MLW+09,*GC12}.   Owing to
the dipole-dipole interaction between  the atoms, they show collective
excitation  of the  form  $|R\rangle =  \frac{1}{\sqrt{N}}\sum_{j=1}^N
|g_1,  g_2, \cdots,r_j,  \cdots g_N\rangle$,  where $|g_k\rangle$  and
$|r_k\rangle$ are the  ground and Rydberg excited states  of the $k$th
atom.  Otherwise  they remain  in the  ground state  $|G\rangle \equiv
|g_1, g_2, \cdots, g_N\rangle$.   The transition between $|g_k\rangle$
and  $|r_k\rangle$ is  tuned  to the  energy of  the  photon used  for
communication.

In the  state $|G\rangle$ the  cloud transitions to  state $|R\rangle$
and  thereby  blocks   the  photon  passage,  whereas   in  the  state
$|R\rangle$ it is transparent to  the photon. The superposition can be
controlled  by  an atom  located  in  a neighboring  potential,  which
interacts  with  the cloud  via  long-range  dipole forces  such  that
depending on  whether it is  in state $|g\rangle$ or  $|r\rangle$, the
cloud is put in the  state $|G\rangle$ and $|R\rangle$.  Preparing the
control  atom  in an  intial  superposition  state then  produces  the
required  superposition of  the Rydberg  cloud acting  as the  quantum
obstacle.

\section{Physical  interpretation   and  conclusions
\label{sec:conclu}}   
Counterfactual communication belies our intuitive expectation that for
Bob  to  send  information  to  Alice, a  physical  particle  must  be
exchanged   between   them.     The   counterfactual   generation   of
entanglement, of the kind presented here, accentuates this puzzle.

One  point  worth  noting  about cryptography  in  the  counterfactual
scenario is  the following: It  might appear that because  no physical
travel happens  in the  open channel during  counterfactual instances,
thus  the information  has  ultimate  security.  But  this  is not  so
\cite{SSS14}, since  Eve's intervention can physicalize  the particle.
Thus security must be judged on the basis of a traditional analysis of
observed  visibilities, photon  counts, details  of an  eavesdropper's
attack, etc.  All the same,  the issue of usefulness of counterfactual
communication for  cryptography and long-distance communication  is an
area worth investigating.

From  a   quantum  foundations  perspective,   counterfactual  quantum
communication appears to  support an ontic interpretation  of the wave
function   that  is   independent   of   any  particular   ontological
framework. During a  $D_1$ detection, there was no  physical travel of
the  particle, in  that,  a  physical travel  would  have resulted  in
absorption.   Yet Bob's  choice  does  influence Alice's  observation,
given  that he  communicates  information. In  order  to maintain  the
philosophical  viewpoint   that  Bob's   action  is  related   to  her
observation  not  by  a  remote influence  but  through  a  continuous
movement of some cause or information in physical space \cite{Vaid14},
we are led to  ascribe some sort of reality to  the wave function (the
vacuum state,  cf.  \cite{Gis13}) that  propagates from Alice  to Bob,
and back  to her.   This conclusion does  not require  any ontological
framework   \cite{HS10},  but   instead   requires  only   operational
considerations about communication (cf. \cite{SS2}).

Here we  summarize how our  scheme for counterfactual  distribution of
entanglement  (and  for  counterfactual quantum  information  transfer
built  on top  of  it)  is distinct  from  the counterfactual  schemes
proposed in  Refs.  \cite{S14a,GCC+14}.  The scheme  in \cite{S14a} is
one for deterministic counterfactual quantum state transfer, requiring
no   classical  communication   and   that   in  \cite{GCC+14}   first
deterministically  and  counterfactually  distributes  state-dependent
entanglement,  which  can  be  used for  deterministic  quantum  state
transfer  with a  1-bit classical  communication (Quantum  information
transfer  of   lower  fidelity  is  possible   without  the  classical
assistance).   In  contrast  to  these  two  schemes,  our  scheme  is
probabilistic. It bears a similarity to the scheme of \cite{GCC+14} in
that   it   distributes   state-dependent  entanglement,   which   can
subequently be used for deterministic quantum state transfer using one
bit  of  classical communication.   The  probabilistic  nature of  our
scheme means (as discussed in  Section \ref{sec:qst}) that on average,
the  minimum  number  of   photons  required  for  the  counterfactual
generation  of state-dependent  entanglement  is $C^{\rm  min}_q =  8$
particles, with  beam splitter  reflectivity $R=\half$.   Further, the
minimum  average  classical   communication  cost  for  counterfactual
quantum  state  transfer is  $1+C^{\rm  min}\approx  4.85$ bits,  with
reflectivity   $R=\sqrt{2}-1\approx0.414$.    From   an   experimental
perspective, of significance is that the schemes of \cite{GCC+14,S14a}
are based on CQZE, which requires  chaining the actions of a number of
beam  splitters.  whereas  our scheme  is probabilistic  and uses  the
experimentally simple Michelsen interferometer set-up.

Our  method   of  generalizing  the  counterfactual   distribution  of
bipartite  entanglement to  multi-partite cat  states, essentially  by
making  the quantum  obstacle  in the  N09 system  to  act jointly  on
multiple particles, can also be applied  to CQZE systems.  To see this
suppose   that   $N$  copies   of   the   $L$-cycle  system   in   Eq.
(\ref{eq:kwiat})  being vertically  stacked, with  the obstacle  being
applied jointly  to all $N$  stack layers.  Then  Eq. (\ref{eq:kwiat})
becomes:
\begin{eqnarray}
|{\rm block}\rangle|0\rangle^{\otimes N} &\longrightarrow& \cos^{LN}\theta
  |{\rm block}\rangle|0\rangle^{\otimes N}  \\
|{\rm pass}\rangle|0\rangle^{\otimes N} &\longrightarrow& 
  |{\rm pass}\rangle(\cos (L\theta)|0\rangle + \sin(L\theta)|1\rangle)^{\otimes N},\nonumber
\label{eq:kwiat++}
\end{eqnarray}
As before let $\theta=\pi/2L$  and $L\rightarrow \infty$.  Then, Bob's
obstacle  in   the  state   $\alpha|{\rm  pass}\rangle   +  \beta|{\rm
  block}\rangle$ leads to:
\begin{eqnarray} 
&& (\alpha|{\rm  pass}\rangle +  \beta|{\rm
  block}\rangle)|0\rangle^{\otimes N} \nonumber\\
&&\hspace{1.0cm}\longrightarrow 
\alpha|{\rm pass}\rangle|1\rangle^{\otimes N}
+ \beta|{\rm block}\rangle|0\rangle^{\otimes N},
\label{eq:cgate++}
\end{eqnarray}
an  $(N+1)$-particle  cat  state,   which  generalizes  the  bipartite
superposition of Eq. (\ref{eq:cgate}). By  placing the above system in
an  external $M$-chain  of interferometers,  so that  each of  the $N$
layers  in  the above  vertical  stack  is  a CQZE  $L$-in-$M$  cyclic
interferometer,   one  can   analogously  implement   a  deterministic
counterfactual distribution of  a $(N+1)$-cat state in  place of state
(\ref{eq:fin-}), that extends the bi-partite scheme of \cite{GCC+14}.

Finally, we  briefly mention some  future directions opened up  by our
work. Our study,  which considers the noiseless case,  may be extended
to the open system situation, which  would be relevant for purposes of
practical implementation.  Multipartite quantum information processing
protocols based on conventional methods of entanglement generation can
be  readily adapted  to  ones based  on  the present  counterfactually
generation entanglement.

\begin{acknowledgments}
RS acknowledges support from the DST project SR/S2/LOP-02/2012.
\end{acknowledgments}

\bibliography{axta}

\begin{thebibliography}{35}%
\makeatletter
\providecommand \@ifxundefined [1]{%
 \@ifx{#1\undefined}
}%
\providecommand \@ifnum [1]{%
 \ifnum #1\expandafter \@firstoftwo
 \else \expandafter \@secondoftwo
 \fi
}%
\providecommand \@ifx [1]{%
 \ifx #1\expandafter \@firstoftwo
 \else \expandafter \@secondoftwo
 \fi
}%
\providecommand \natexlab [1]{#1}%
\providecommand \enquote  [1]{``#1''}%
\providecommand \bibnamefont  [1]{#1}%
\providecommand \bibfnamefont [1]{#1}%
\providecommand \citenamefont [1]{#1}%
\providecommand \href@noop [0]{\@secondoftwo}%
\providecommand \href [0]{\begingroup \@sanitize@url \@href}%
\providecommand \@href[1]{\@@startlink{#1}\@@href}%
\providecommand \@@href[1]{\endgroup#1\@@endlink}%
\providecommand \@sanitize@url [0]{\catcode `\\12\catcode `\$12\catcode
  `\&12\catcode `\#12\catcode `\^12\catcode `\_12\catcode `\%12\relax}%
\providecommand \@@startlink[1]{}%
\providecommand \@@endlink[0]{}%
\providecommand \url  [0]{\begingroup\@sanitize@url \@url }%
\providecommand \@url [1]{\endgroup\@href {#1}{\urlprefix }}%
\providecommand \urlprefix  [0]{URL }%
\providecommand \Eprint [0]{\href }%
\providecommand \doibase [0]{http://dx.doi.org/}%
\providecommand \selectlanguage [0]{\@gobble}%
\providecommand \bibinfo  [0]{\@secondoftwo}%
\providecommand \bibfield  [0]{\@secondoftwo}%
\providecommand \translation [1]{[#1]}%
\providecommand \BibitemOpen [0]{}%
\providecommand \bibitemStop [0]{}%
\providecommand \bibitemNoStop [0]{.\EOS\space}%
\providecommand \EOS [0]{\spacefactor3000\relax}%
\providecommand \BibitemShut  [1]{\csname bibitem#1\endcsname}%
\let\auto@bib@innerbib\@empty
\bibitem [{\citenamefont {Jozsa}(1998)}]{Joz98}%
  \BibitemOpen
  \bibfield  {author} {\bibinfo {author} {\bibfnamefont {R.}~\bibnamefont
  {Jozsa}},\ }in\ \href@noop {} {\textit {\bibinfo {booktitle} {Lecture Notes in
  Computer Science}}},\ Vol.\ \bibinfo {volume} {1509},\ \bibinfo {editor}
  {edited by\ \bibinfo {editor} {\bibfnamefont {C.~P.}\ \bibnamefont
  {Williams}}}\ (\bibinfo  {publisher} {Springer},\ \bibinfo {year} {1998})\
  p.\ \bibinfo {pages} {103}\BibitemShut {NoStop}%
\bibitem [{\citenamefont {Mitchison}\ and\ \citenamefont {Jozsa}(2001)}]{MJ01}%
  \BibitemOpen
  \bibfield  {author} {\bibinfo {author} {\bibfnamefont {G.}~\bibnamefont
  {Mitchison}}\ and\ \bibinfo {author} {\bibfnamefont {R.}~\bibnamefont
  {Jozsa}},\ }\bibfield  {title} {\enquote {\bibinfo {title} {Counterfactual
  computation},}\ }\href@noop {} {\bibfield  {journal} {\bibinfo  {journal}
  {Proc. Roy. Soc. Lond. A457}\ ,\ \bibinfo {pages} {1175--1194}} (\bibinfo
  {year} {2001})}\BibitemShut {NoStop}%
\bibitem [{\citenamefont {Elitzur}\ and\ \citenamefont {Vaidman}(1993)}]{EV93}%
  \BibitemOpen
  \bibfield  {author} {\bibinfo {author} {\bibfnamefont {A.~C.}\ \bibnamefont
  {Elitzur}}\ and\ \bibinfo {author} {\bibfnamefont {L.}~\bibnamefont
  {Vaidman}},\ }\bibfield  {title} {\enquote {\bibinfo {title}
  {Interaction-free measurements},}\ }\href@noop {} {\bibfield  {journal}
  {\bibinfo  {journal} {Found. of Phys.}\ }\textbf {\bibinfo {volume} {23}},\
  \bibinfo {pages} {987} (\bibinfo {year} {1993})}\BibitemShut {NoStop}%
\bibitem [{\citenamefont {Kwiat}\ \textit {et~al.}(1995)\citenamefont {Kwiat},
  \citenamefont {Weinfurter}, \citenamefont {Herzog}, \citenamefont
  {Zeilinger},\ and\ \citenamefont {Kasevich}}]{KWH+95}%
  \BibitemOpen
  \bibfield  {author} {\bibinfo {author} {\bibfnamefont {P.~G.}\ \bibnamefont
  {Kwiat}}, \bibinfo {author} {\bibfnamefont {H.}~\bibnamefont {Weinfurter}},
  \bibinfo {author} {\bibfnamefont {T.}~\bibnamefont {Herzog}}, \bibinfo
  {author} {\bibfnamefont {A.}~\bibnamefont {Zeilinger}}, \ and\ \bibinfo
  {author} {\bibfnamefont {M.~A.}\ \bibnamefont {Kasevich}},\ }\bibfield
  {title} {\enquote {\bibinfo {title} {Interaction-free measurement},}\
  }\href@noop {} {\bibfield  {journal} {\bibinfo  {journal} {Phys. Rev. Lett.}\
  }\textbf {\bibinfo {volume} {74}},\ \bibinfo {pages} {4763--4766} (\bibinfo
  {year} {1995})}\BibitemShut {NoStop}%
\bibitem [{\citenamefont {Guo}\ and\ \citenamefont {Shi}(1999)}]{GS99}%
  \BibitemOpen
  \bibfield  {author} {\bibinfo {author} {\bibfnamefont {G.-C.}\ \bibnamefont
  {Guo}}\ and\ \bibinfo {author} {\bibfnamefont {B.-S.}\ \bibnamefont {Shi}},\
  }\bibfield  {title} {\enquote {\bibinfo {title} {Quantum cryptography based
  on interaction-free measurement},}\ }\href@noop {} {\bibfield  {journal}
  {\bibinfo  {journal} {Phys. Lett. A}\ }\textbf {\bibinfo {volume} {256}},\
  \bibinfo {pages} {109} (\bibinfo {year} {1999})}\BibitemShut {NoStop}%
\bibitem [{\citenamefont {Hosten}\ \textit {et~al.}(2006)\citenamefont {Hosten},
  \citenamefont {Rakher}, \citenamefont {Barreiro}, \citenamefont {Peters},\
  and\ \citenamefont {Kwiat}}]{HRB+06}%
  \BibitemOpen
  \bibfield  {author} {\bibinfo {author} {\bibfnamefont {O.}~\bibnamefont
  {Hosten}}, \bibinfo {author} {\bibfnamefont {M.~T.}\ \bibnamefont {Rakher}},
  \bibinfo {author} {\bibfnamefont {J.~T.}\ \bibnamefont {Barreiro}}, \bibinfo
  {author} {\bibfnamefont {N.~A.}\ \bibnamefont {Peters}}, \ and\ \bibinfo
  {author} {\bibfnamefont {P.~G.}\ \bibnamefont {Kwiat}},\ }\bibfield  {title}
  {\enquote {\bibinfo {title} {Counterfactual quantum computation through
  quantum interrogation},}\ }\href@noop {} {\bibfield  {journal} {\bibinfo
  {journal} {Nature}\ }\textbf {\bibinfo {volume} {439}},\ \bibinfo {pages}
  {949--952} (\bibinfo {year} {2006})}\BibitemShut {NoStop}%
\bibitem [{\citenamefont {Vaidman}(2007)}]{vai07}%
  \BibitemOpen
  \bibfield  {author} {\bibinfo {author} {\bibfnamefont {L.}~\bibnamefont
  {Vaidman}},\ }\bibfield  {title} {\enquote {\bibinfo {title} {Impossibility
  of the counterfactual computation for all possible outcomes},}\ }\href
  {\doibase 10.1103/PhysRevLett.98.160403} {\bibfield  {journal} {\bibinfo
  {journal} {Phys. Rev. Lett.}\ }\textbf {\bibinfo {volume} {98}},\ \bibinfo
  {pages} {160403} (\bibinfo {year} {2007})}\BibitemShut {NoStop}%
\bibitem [{\citenamefont {Noh}(2009)}]{N09}%
  \BibitemOpen
  \bibfield  {author} {\bibinfo {author} {\bibfnamefont {T.-G.}\ \bibnamefont
  {Noh}},\ }\bibfield  {title} {\enquote {\bibinfo {title} {Counterfactual
  quantum cryptography},}\ }\href {\doibase 10.1103/PhysRevLett.103.230501}
  {\bibfield  {journal} {\bibinfo  {journal} {Phys. Rev. Lett.}\ }\textbf
  {\bibinfo {volume} {103}},\ \bibinfo {pages} {230501} (\bibinfo {year}
  {2009})}\BibitemShut {NoStop}%
\bibitem [{\citenamefont {Sun}\ and\ \citenamefont {Wen}(2010)}]{SW10}%
  \BibitemOpen
  \bibfield  {author} {\bibinfo {author} {\bibfnamefont {Y.}~\bibnamefont
  {Sun}}\ and\ \bibinfo {author} {\bibfnamefont {Q.-Y.}\ \bibnamefont {Wen}},\
  }\bibfield  {title} {\enquote {\bibinfo {title} {Counterfactual quantum key
  distribution with high efficiency},}\ }\href@noop {} {\bibfield  {journal}
  {\bibinfo  {journal} {Phys. Rev. A}\ }\textbf {\bibinfo {volume} {82}},\
  \bibinfo {pages} {052318} (\bibinfo {year} {2010})}\BibitemShut {NoStop}%
\bibitem [{\citenamefont {Zhang}\ \textit {et~al.}()\citenamefont {Zhang},
  \citenamefont {Liu},\ and\ \citenamefont {Zhang}}]{ZLZ14}%
  \BibitemOpen
  \bibfield  {author} {\bibinfo {author} {\bibfnamefont {S.}~\bibnamefont
  {Zhang}}, \bibinfo {author} {\bibfnamefont {X.-T.}\ \bibnamefont {Liu}}, \
  and\ \bibinfo {author} {\bibfnamefont {B.}~\bibnamefont {Zhang}},\
  }\href@noop {} {\enquote {\bibinfo {title} {Direct counterfactual quantum
  communication with high efficiency},}\ }\bibinfo {note}
  {ArXiv:1410.2769}\BibitemShut {NoStop}%
\bibitem [{\citenamefont {Yin}\ \textit {et~al.}(2010)\citenamefont {Yin},
  \citenamefont {Li}, \citenamefont {Chen}, \citenamefont {Han},\ and\
  \citenamefont {Guo}}]{YLC+10}%
  \BibitemOpen
  \bibfield  {author} {\bibinfo {author} {\bibfnamefont {Z.-Q.}\ \bibnamefont
  {Yin}}, \bibinfo {author} {\bibfnamefont {H.-W.}\ \bibnamefont {Li}},
  \bibinfo {author} {\bibfnamefont {W.}~\bibnamefont {Chen}}, \bibinfo {author}
  {\bibfnamefont {Z.-F.}\ \bibnamefont {Han}}, \ and\ \bibinfo {author}
  {\bibfnamefont {G.-C.}\ \bibnamefont {Guo}},\ }\bibfield  {title} {\enquote
  {\bibinfo {title} {Security of counterfactual quantum cryptography},}\
  }\href@noop {} {\bibfield  {journal} {\bibinfo  {journal} {Phys. Rev. A}\
  }\textbf {\bibinfo {volume} {82}},\ \bibinfo {pages} {042335} (\bibinfo
  {year} {2010})}\BibitemShut {NoStop}%
\bibitem [{\citenamefont {Zhang}\ \textit
  {et~al.}(2012{\natexlab{a}})\citenamefont {Zhang}, \citenamefont {Wang},
  \citenamefont {Tang},\ and\ \citenamefont {Zhang}}]{ZWT+12}%
  \BibitemOpen
  \bibfield  {author} {\bibinfo {author} {\bibfnamefont {S.}~\bibnamefont
  {Zhang}}, \bibinfo {author} {\bibfnamefont {J.}~\bibnamefont {Wang}},
  \bibinfo {author} {\bibfnamefont {C.-J.}\ \bibnamefont {Tang}}, \ and\
  \bibinfo {author} {\bibfnamefont {Q.}~\bibnamefont {Zhang}},\ }\bibfield
  {title} {\enquote {\bibinfo {title} {Security proof of counterfactual quantum
  cryptography against general intercept-resend attacks and its
  vulnerability},}\ }\href@noop {} {\bibfield  {journal} {\bibinfo  {journal}
  {Chin. Phys. B}\ }\textbf {\bibinfo {volume} {21}},\ \bibinfo {pages}
  {060303} (\bibinfo {year} {2012}{\natexlab{a}})}\BibitemShut {NoStop}%
\bibitem [{\citenamefont {Zhang}\ \textit
  {et~al.}(2012{\natexlab{b}})\citenamefont {Zhang}, \citenamefont {Wang},\
  and\ \citenamefont {Tang}}]{ZWT12}%
  \BibitemOpen
  \bibfield  {author} {\bibinfo {author} {\bibfnamefont {S.}~\bibnamefont
  {Zhang}}, \bibinfo {author} {\bibfnamefont {J.}~\bibnamefont {Wang}}, \ and\
  \bibinfo {author} {\bibfnamefont {C.~J.}\ \bibnamefont {Tang}},\ }\bibfield
  {title} {\enquote {\bibinfo {title} {Counterfactual attack on counterfactual
  quantum key distribution},}\ }\href@noop {} {\bibfield  {journal} {\bibinfo
  {journal} {Europhys. Lett.}\ }\textbf {\bibinfo {volume} {98}},\ \bibinfo
  {pages} {30012} (\bibinfo {year} {2012}{\natexlab{b}})}\BibitemShut {NoStop}%
\bibitem [{\citenamefont {Salih}\ \textit {et~al.}(2013)\citenamefont {Salih},
  \citenamefont {Li}, \citenamefont {Al-Amri},\ and\ \citenamefont
  {Zubairy}}]{SLA+13}%
  \BibitemOpen
  \bibfield  {author} {\bibinfo {author} {\bibfnamefont {H.}~\bibnamefont
  {Salih}}, \bibinfo {author} {\bibfnamefont {Z.-H.}\ \bibnamefont {Li}},
  \bibinfo {author} {\bibfnamefont {M.}~\bibnamefont {Al-Amri}}, \ and\
  \bibinfo {author} {\bibfnamefont {M.~S.}\ \bibnamefont {Zubairy}},\
  }\bibfield  {title} {\enquote {\bibinfo {title} {Protocol for direct
  counterfactual quantum communication},}\ }\href {\doibase
  10.1103/PhysRevLett.110.170502} {\bibfield  {journal} {\bibinfo  {journal}
  {Phys. Rev. Lett.}\ }\textbf {\bibinfo {volume} {110}},\ \bibinfo {pages}
  {170502} (\bibinfo {year} {2013})}\BibitemShut {NoStop}%
\bibitem [{\citenamefont {Shenoy~H.}\ \textit {et~al.}(2013)\citenamefont
  {Shenoy~H.}, \citenamefont {Srikanth},\ and\ \citenamefont
  {Srinivas}}]{SSS1}%
  \BibitemOpen
  \bibfield  {author} {\bibinfo {author} {\bibfnamefont {A.}~\bibnamefont
  {Shenoy~H.}}, \bibinfo {author} {\bibfnamefont {R.}~\bibnamefont {Srikanth}},
  \ and\ \bibinfo {author} {\bibfnamefont {T.}~\bibnamefont {Srinivas}},\
  }\bibfield  {title} {\enquote {\bibinfo {title} {Semi-counterfactual
  cryptography},}\ }\href@noop {} {\bibfield  {journal} {\bibinfo  {journal}
  {Europhys. Lett.}\ }\textbf {\bibinfo {volume} {103}},\ \bibinfo {pages}
  {60008} (\bibinfo {year} {2013})}\BibitemShut {NoStop}%
\bibitem [{\citenamefont {Zhang}\ \textit {et~al.}(2013)\citenamefont {Zhang},
  \citenamefont {Guo}, \citenamefont {Gao}, \citenamefont {Liu},\ and\
  \citenamefont {Wen}}]{ZGG+13}%
  \BibitemOpen
  \bibfield  {author} {\bibinfo {author} {\bibfnamefont {J.-L.}\ \bibnamefont
  {Zhang}}, \bibinfo {author} {\bibfnamefont {F.-Z.}\ \bibnamefont {Guo}},
  \bibinfo {author} {\bibfnamefont {F.}~\bibnamefont {Gao}}, \bibinfo {author}
  {\bibfnamefont {B.}~\bibnamefont {Liu}}, \ and\ \bibinfo {author}
  {\bibfnamefont {Q.-Y.}\ \bibnamefont {Wen}},\ }\bibfield  {title} {\enquote
  {\bibinfo {title} {Private database queries based on counterfactual quantum
  key distribution},}\ }\href@noop {} {\bibfield  {journal} {\bibinfo
  {journal} {Phys. Rev. A}\ }\textbf {\bibinfo {volume} {88}},\ \bibinfo
  {pages} {022334} (\bibinfo {year} {2013})}\BibitemShut {NoStop}%
\bibitem [{\citenamefont {Shenoy~H.}\ \textit {et~al.}(2014)\citenamefont
  {Shenoy~H.}, \citenamefont {Srikanth},\ and\ \citenamefont
  {Srinivas}}]{SSS14}%
  \BibitemOpen
  \bibfield  {author} {\bibinfo {author} {\bibfnamefont {A.}~\bibnamefont
  {Shenoy~H.}}, \bibinfo {author} {\bibfnamefont {R.}~\bibnamefont {Srikanth}},
  \ and\ \bibinfo {author} {\bibfnamefont {T.}~\bibnamefont {Srinivas}},\
  }\bibfield  {title} {\enquote {\bibinfo {title} {Counterfactual quantum
  certificate authorization},}\ }\href {\doibase 10.1103/PhysRevA.89.052307}
  {\bibfield  {journal} {\bibinfo  {journal} {Phys. Rev. A}\ }\textbf {\bibinfo
  {volume} {89}},\ \bibinfo {pages} {052307} (\bibinfo {year}
  {2014})}\BibitemShut {NoStop}%
\bibitem [{\citenamefont {Salih}(2014)}]{Sal14}%
  \BibitemOpen
  \bibfield  {author} {\bibinfo {author} {\bibfnamefont {H.}~\bibnamefont
  {Salih}},\ }\bibfield  {title} {\enquote {\bibinfo {title} {Tripartite
  counterfactual quantum cryptography},}\ }\href@noop {} {\bibfield  {journal}
  {\bibinfo  {journal} {Phys. Rev. A}\ }\textbf {\bibinfo {volume} {90}},\
  \bibinfo {pages} {012333} (\bibinfo {year} {2014})}\BibitemShut {NoStop}%
\bibitem [{\citenamefont {Yin}\ \textit {et~al.}(2012)\citenamefont {Yin},
  \citenamefont {Li}, \citenamefont {Yao}, \citenamefont {Zhang}, \citenamefont
  {W.}, \citenamefont {Chen}, \citenamefont {Guo},\ and\ \citenamefont
  {Han}}]{YLY+12}%
  \BibitemOpen
  \bibfield  {author} {\bibinfo {author} {\bibfnamefont {Z.-Q.}\ \bibnamefont
  {Yin}}, \bibinfo {author} {\bibfnamefont {H.-W.}\ \bibnamefont {Li}},
  \bibinfo {author} {\bibfnamefont {Y.}~\bibnamefont {Yao}}, \bibinfo {author}
  {\bibfnamefont {C.-M.}\ \bibnamefont {Zhang}}, \bibinfo {author}
  {\bibfnamefont {S.}~\bibnamefont {W.}}, \bibinfo {author} {\bibfnamefont
  {W.}~\bibnamefont {Chen}}, \bibinfo {author} {\bibfnamefont {G.-C.}\
  \bibnamefont {Guo}}, \ and\ \bibinfo {author} {\bibfnamefont {Z.-F.}\
  \bibnamefont {Han}},\ }\bibfield  {title} {\enquote {\bibinfo {title}
  {Counterfactual quantum cryptography based on weak coherent states},}\ }\href
  {\doibase 10.1103/PhysRevA.86.022313} {\bibfield  {journal} {\bibinfo
  {journal} {Phys. Rev. A}\ }\textbf {\bibinfo {volume} {86}},\ \bibinfo
  {pages} {022313} (\bibinfo {year} {2012})}\BibitemShut {NoStop}%
\bibitem [{\citenamefont {Brida}\ \textit {et~al.}(2012)\citenamefont {Brida},
  \citenamefont {Cavanna}, \citenamefont {Degiovanni}, \citenamefont
  {Genovese},\ and\ \citenamefont {Traina}}]{BCD+12}%
  \BibitemOpen
  \bibfield  {author} {\bibinfo {author} {\bibfnamefont {G.}~\bibnamefont
  {Brida}}, \bibinfo {author} {\bibfnamefont {A.}~\bibnamefont {Cavanna}},
  \bibinfo {author} {\bibfnamefont {I.~P.}\ \bibnamefont {Degiovanni}},
  \bibinfo {author} {\bibfnamefont {M.}~\bibnamefont {Genovese}}, \ and\
  \bibinfo {author} {\bibfnamefont {P.}~\bibnamefont {Traina}},\ }\bibfield
  {title} {\enquote {\bibinfo {title} {Experimental realization of
  counterfactual quantum cryptography},}\ }\href@noop {} {\bibfield  {journal}
  {\bibinfo  {journal} {Laser Phys. Lett.}\ }\textbf {\bibinfo {volume} {9}},\
  \bibinfo {pages} {247} (\bibinfo {year} {2012})}\BibitemShut {NoStop}%
\bibitem [{\citenamefont {Liu}\ \textit {et~al.}(2012)\citenamefont {Liu},
  \citenamefont {Ju}, \citenamefont {Liang}, \citenamefont {Tang},
  \citenamefont {Tu}, \citenamefont {Zhou}, \citenamefont {Peng}, \citenamefont
  {Chen}, \citenamefont {Chen}, \citenamefont {Chen},\ and\ \citenamefont
  {Pan}}]{LJL+12}%
  \BibitemOpen
  \bibfield  {author} {\bibinfo {author} {\bibfnamefont {Y.}~\bibnamefont
  {Liu}}, \bibinfo {author} {\bibfnamefont {L.}~\bibnamefont {Ju}}, \bibinfo
  {author} {\bibfnamefont {X.-L.}\ \bibnamefont {Liang}}, \bibinfo {author}
  {\bibfnamefont {S.-B.}\ \bibnamefont {Tang}}, \bibinfo {author}
  {\bibfnamefont {G.-L.~S.}\ \bibnamefont {Tu}}, \bibinfo {author}
  {\bibfnamefont {L.}~\bibnamefont {Zhou}}, \bibinfo {author} {\bibfnamefont
  {C.-Z.}\ \bibnamefont {Peng}}, \bibinfo {author} {\bibfnamefont
  {K.}~\bibnamefont {Chen}}, \bibinfo {author} {\bibfnamefont {T.-Y.}\
  \bibnamefont {Chen}}, \bibinfo {author} {\bibfnamefont {Z.-B.}\ \bibnamefont
  {Chen}}, \ and\ \bibinfo {author} {\bibfnamefont {J.-W.}\ \bibnamefont
  {Pan}},\ }\bibfield  {title} {\enquote {\bibinfo {title} {Experimental
  demonstration of counterfactual quantum communication},}\ }\href@noop {}
  {\bibfield  {journal} {\bibinfo  {journal} {Phys. Rev. Lett.}\ }\textbf
  {\bibinfo {volume} {109}},\ \bibinfo {pages} {030501} (\bibinfo {year}
  {2012})}\BibitemShut {NoStop}%
\bibitem [{\citenamefont {Vaidman}(2014)}]{vai14}%
  \BibitemOpen
  \bibfield  {author} {\bibinfo {author} {\bibfnamefont {L.}~\bibnamefont
  {Vaidman}},\ }\bibfield  {title} {\enquote {\bibinfo {title} {Comment on
  `{P}rotocol for direct counterfactual quantum communication'},}\ }\href
  {\doibase 10.1103/PhysRevLett.112.208901} {\bibfield  {journal} {\bibinfo
  {journal} {Phys. Rev. Lett.}\ }\textbf {\bibinfo {volume} {112}},\ \bibinfo
  {pages} {208901} (\bibinfo {year} {2014})}\BibitemShut {NoStop}%
\bibitem [{\citenamefont {Salih}\ \textit {et~al.}(2014)\citenamefont {Salih},
  \citenamefont {Li}, \citenamefont {Al-Amri},\ and\ \citenamefont
  {Zubairy}}]{SLA+14}%
  \BibitemOpen
  \bibfield  {author} {\bibinfo {author} {\bibfnamefont {H.}~\bibnamefont
  {Salih}}, \bibinfo {author} {\bibfnamefont {Z.-Hong}\ \bibnamefont {Li}},
  \bibinfo {author} {\bibfnamefont {M.}~\bibnamefont {Al-Amri}}, \ and\
  \bibinfo {author} {\bibfnamefont {M.~S.}\ \bibnamefont {Zubairy}},\
  }\bibfield  {title} {\enquote {\bibinfo {title} {Salih et~al. reply:},}\
  }\href {\doibase 10.1103/PhysRevLett.112.208902} {\bibfield  {journal}
  {\bibinfo  {journal} {Phys. Rev. Lett.}\ }\textbf {\bibinfo {volume} {112}},\
  \bibinfo {pages} {208902} (\bibinfo {year} {2014})}\BibitemShut {NoStop}%
\bibitem [{\citenamefont {Gisin}(2013)}]{Gis13}%
  \BibitemOpen
  \bibfield  {author} {\bibinfo {author} {\bibfnamefont {N.}~\bibnamefont
  {Gisin}},\ }\bibfield  {title} {\enquote {\bibinfo {title} {Optical
  communication without photons},}\ }\href@noop {} {\bibfield  {journal}
  {\bibinfo  {journal} {Phys. Rev. A}\ }\textbf {\bibinfo {volume} {88}},\
  \bibinfo {pages} {030301} (\bibinfo {year} {2013})}\BibitemShut {NoStop}%
\bibitem [{\citenamefont {Vaidman}()}]{Vaid14}%
  \BibitemOpen
  \bibfield  {author} {\bibinfo {author} {\bibfnamefont {L.}~\bibnamefont
  {Vaidman}},\ }\href@noop {} {\enquote {\bibinfo {title} {Counterfactaulity of
  counterfactual communication},}\ }\bibinfo {note}
  {ArXiv:1410.2723}\BibitemShut {NoStop}%
\bibitem [{\citenamefont {Salih}()}]{S14a}%
  \BibitemOpen
  \bibfield  {author} {\bibinfo {author} {\bibfnamefont {H.}~\bibnamefont
  {Salih}},\ }\href@noop {} {\enquote {\bibinfo {title} {Protocol for
  counterfactually transporting an unknown qubit},}\ }\bibinfo {note}
  {ArXiv:1404.2200}\BibitemShut {NoStop}%
\bibitem [{\citenamefont {Guo}\ \textit {et~al.}(2014)\citenamefont {Guo},
  \citenamefont {Cheng}, \citenamefont {Chen}, \citenamefont {Wang},\ and\
  \citenamefont {Zhang}}]{GCC+14}%
  \BibitemOpen
  \bibfield  {author} {\bibinfo {author} {\bibfnamefont {Q.}~\bibnamefont
  {Guo}}, \bibinfo {author} {\bibfnamefont {L.-Y.}\ \bibnamefont {Cheng}},
  \bibinfo {author} {\bibfnamefont {L.}~\bibnamefont {Chen}}, \bibinfo {author}
  {\bibfnamefont {H.-F.}\ \bibnamefont {Wang}}, \ and\ \bibinfo {author}
  {\bibfnamefont {S.}~\bibnamefont {Zhang}},\ }\bibfield  {title} {\enquote
  {\bibinfo {title} {Counterfactual quantum-information transfer without
  transmitting any physical particles},}\ }\href@noop {} {\bibfield  {journal}
  {\bibinfo  {journal} {Sci. Rep.}\ }\textbf {\bibinfo {volume} {5}},\ \bibinfo
  {pages} {8416} (\bibinfo {year} {2014})}\BibitemShut {NoStop}%
\bibitem [{\citenamefont {Nielsen}\ and\ \citenamefont {Chuang}(2000)}]{NC00}%
  \BibitemOpen
  \bibfield  {author} {\bibinfo {author} {\bibfnamefont {M.}~\bibnamefont
  {Nielsen}}\ and\ \bibinfo {author} {\bibfnamefont {I.}~\bibnamefont
  {Chuang}},\ }\href@noop {} {\textit {\bibinfo {title} {Quantum computation and
  quantum information}}}\ (\bibinfo  {publisher} {Cambridge},\ \bibinfo {year}
  {2000})\BibitemShut {NoStop}%
\bibitem [{\citenamefont {Wineland}(2013)}]{W13}%
  \BibitemOpen
  \bibfield  {author} {\bibinfo {author} {\bibfnamefont {D.}~\bibnamefont
  {Wineland}},\ }\bibfield  {title} {\enquote {\bibinfo {title} {Superposition,
  entanglement, and raising schrodinger’s cat},}\ }\href@noop {} {\bibfield
  {journal} {\bibinfo  {journal} {Rev. Mod. Phys.}\ }\textbf {\bibinfo {volume}
  {85}},\ \bibinfo {pages} {1103--1114} (\bibinfo {year} {2013})},\ \bibinfo
  {note} {nobel lecture}\BibitemShut {NoStop}%
\bibitem [{\citenamefont {Gerlich}\ \textit {et~al.}(2011)\citenamefont {Gerlich}
  \textit {et~al.}}]{G+11}%
  \BibitemOpen
  \bibfield  {author} {\bibinfo {author} {\bibfnamefont {S.}~\bibnamefont
  {Gerlich}} \textit {et~al.},\ }\href@noop {} {\bibfield  {journal} {\bibinfo
  {journal} {Nature Comm.}\ }\textbf {\bibinfo {volume} {2}},\ \bibinfo {pages}
  {263} (\bibinfo {year} {2011})}\BibitemShut {NoStop}%
\bibitem [{\citenamefont {Kwiat}\ \textit {et~al.}(1999)\citenamefont {Kwiat},
  \citenamefont {White}, \citenamefont {Mitchell}, \citenamefont {Nairz},
  \citenamefont {Weihs}, \citenamefont {Weinfurter},\ and\ \citenamefont
  {Zeilinger}}]{KWM+99}%
  \BibitemOpen
  \bibfield  {author} {\bibinfo {author} {\bibfnamefont {P.~G.}\ \bibnamefont
  {Kwiat}}, \bibinfo {author} {\bibfnamefont {A.~G.}\ \bibnamefont {White}},
  \bibinfo {author} {\bibfnamefont {J.~R.}\ \bibnamefont {Mitchell}}, \bibinfo
  {author} {\bibfnamefont {O.}~\bibnamefont {Nairz}}, \bibinfo {author}
  {\bibfnamefont {G.}~\bibnamefont {Weihs}}, \bibinfo {author} {\bibfnamefont
  {H.}~\bibnamefont {Weinfurter}}, \ and\ \bibinfo {author} {\bibfnamefont
  {A.}~\bibnamefont {Zeilinger}},\ }\bibfield  {title} {\enquote {\bibinfo
  {title} {High-efficiency quantum interrogation measurements via the quantum
  zeno effect},}\ }\href {\doibase 10.1103/PhysRevLett.83.4725} {\bibfield
  {journal} {\bibinfo  {journal} {Phys. Rev. Lett.}\ }\textbf {\bibinfo
  {volume} {83}},\ \bibinfo {pages} {4725--4728} (\bibinfo {year}
  {1999})}\BibitemShut {NoStop}%
\bibitem [{\citenamefont {M\"uller}\ \textit {et~al.}(2009)\citenamefont
  {M\"uller}, \citenamefont {Lesanovsky}, \citenamefont {Weimer}, \citenamefont
  {B\"uchler}, \citenamefont {P.},\ and\ \citenamefont {Zoller}}]{MLW+09}%
  \BibitemOpen
  \bibfield  {author} {\bibinfo {author} {\bibfnamefont {M.}~\bibnamefont
  {M\"uller}}, \bibinfo {author} {\bibfnamefont {I.}~\bibnamefont
  {Lesanovsky}}, \bibinfo {author} {\bibfnamefont {H.}~\bibnamefont {Weimer}},
  \bibinfo {author} {\bibnamefont {B\"uchler}}, \bibinfo {author}
  {\bibfnamefont {H.}~\bibnamefont {P.}}, \ and\ \bibinfo {author}
  {\bibfnamefont {P.}~\bibnamefont {Zoller}},\ }\bibfield  {title} {\enquote
  {\bibinfo {title} {Mesoscopic rydberg gate based on electromagnetically
  induced transparency},}\ }\href@noop {} {\bibfield  {journal} {\bibinfo
  {journal} {Phys. Rev. Lett.}\ }\textbf {\bibinfo {volume} {102}},\ \bibinfo
  {pages} {170502} (\bibinfo {year} {2009})}\BibitemShut {NoStop}%
\bibitem [{\citenamefont {Garcia-Escartin}\ and\ \citenamefont
  {Chamorro-Posada}(2012)}]{GC12}%
  \BibitemOpen
  \bibfield  {author} {\bibinfo {author} {\bibfnamefont {J.~C.}\ \bibnamefont
  {Garcia-Escartin}}\ and\ \bibinfo {author} {\bibfnamefont {P.}~\bibnamefont
  {Chamorro-Posada}},\ }\bibfield  {title} {\enquote {\bibinfo {title}
  {Counterfactual rydberg gate for photons},}\ }\href@noop {} {\bibfield
  {journal} {\bibinfo  {journal} {Phys. Rev. A}\ }\textbf {\bibinfo {volume}
  {85}},\ \bibinfo {pages} {032309} (\bibinfo {year} {2012})}\BibitemShut
  {NoStop}%
\bibitem [{\citenamefont {Harrigan}\ and\ \citenamefont
  {Spekkens}(2010)}]{HS10}%
  \BibitemOpen
  \bibfield  {author} {\bibinfo {author} {\bibfnamefont {N.}~\bibnamefont
  {Harrigan}}\ and\ \bibinfo {author} {\bibfnamefont {R.~W.}\ \bibnamefont
  {Spekkens}},\ }\bibfield  {title} {\enquote {\bibinfo {title} {Einstein,
  incompleteness, and the epistemic view of quantum states},}\ }\href@noop {}
  {\bibfield  {journal} {\bibinfo  {journal} {Found. Phys.}\ }\textbf {\bibinfo
  {volume} {40}},\ \bibinfo {pages} {125} (\bibinfo {year} {2010})}\BibitemShut
  {NoStop}%
\bibitem [{\citenamefont {Shenoy~H.}\ and\ \citenamefont
  {Srikanth}(2013)}]{SS2}%
  \BibitemOpen
  \bibfield  {author} {\bibinfo {author} {\bibfnamefont {A.}~\bibnamefont
  {Shenoy~H.}}\ and\ \bibinfo {author} {\bibfnamefont {R.}~\bibnamefont
  {Srikanth}},\ }\href@noop {} {\enquote {\bibinfo {title} {The wave-function
  is real but nonphysical: A view from counterfactual quantum cryptography},}\
  } (\bibinfo {year} {2013}),\ \bibinfo {note} {arXiv:1311.7127}\BibitemShut
  {NoStop}%
\end{thebibliography}%

\end{document}